\documentclass[journal]{IEEEtran}

\usepackage{cite}
\usepackage{amsmath,amssymb,amsfonts}
\usepackage{algorithmic}
\usepackage{graphicx}
\usepackage{float}
\usepackage{textcomp}
\usepackage{comment}
\usepackage[dvipsnames]{xcolor}

\def\BibTeX{{\rm B\kern-.05em{\sc i\kern-.025em b}\kern-.08em
    T\kern-.1667em\lower.7ex\hbox{E}\kern-.125emX}}

\begin{document}

\title{A Complex Frequency-Based Control for Inverter-Based Resources}
\author{%
  Rodrigo~Bernal,~\IEEEmembership{IEEE Member,} and
  Federico~Milano,~\IEEEmembership{IEEE Fellow}
  \thanks{This work is supported by Sustainable Energy Authority of Ireland (SEAI) by funding R.~Bernal and F.~Milano under project FRESLIPS, Grant No.~RDD/00681.}%
}
\maketitle

\begin{abstract}
  This paper proposes a novel control for Inverter-based Resources (IBRs) based on the Complex Frequency (CF) concept.  The controller's objective is to maintain a constant CF of the voltage at the terminals of the IBR by adjusting its current reference.  This current is imposed based on the well-known power flow equation, the dynamics of which are calculated through the estimation of the CF of the voltages of the adjacent buses. Performance is evaluated by analyzing local variations in frequency and magnitude of the voltage, as well as the response of the system’s Center of Inertia (CoI) frequency, and then compared with conventional frequency droop, PI voltage controllers and virtual inertia. The case study utilizes the WSCC 9-bus system and a 1479-bus model of the Irish transmission grid and considers various contingencies and sensitivities such as the impact of limiters, delays, noise, R/X ratio, and EMT dynamics.  Results show that the proposed scheme consistently outperforms the conventional controllers, leading to significant improvements in the overall dynamic response of the system.
\end{abstract}

\begin{IEEEkeywords}
Complex frequency (CF), frequency control, inverter-based resource (IBR), voltage control.
\end{IEEEkeywords}

\section{Introduction}
\label{sec:intro}

\subsection{Motivation}
\label{sub:motivation}

Power systems are currently undergoing the replacement of Synchronous Machines (SMs) with inverter-based resources (IBRs).  This transition introduces new system dynamics, characterized by increased speed and technological heterogeneity.
In the initial stages of this transformation, the focus was on adapting converters to the grid.  Synchronization approaches were required and employed to utilize all available energy \cite{converter_sync}.  As converters gained a larger share in the systems, efforts were directed towards technical regulations.  New requirements related to voltage support, frequency regulation, voltage ride-through (VRT), power quality, and active/reactive power control were established to ensure grid stability \cite{converter_requirements_ctrlModes}.  As converters become more dominant, new and more flexible control structures are required.  Nevertheless, the dynamic interaction of these controllers with the rest of the system is yet to be fully understood and is still one of the major challenges currently faced by modern systems \cite{Fundations}.
IBRs are essential in high-renewable systems, isolated microgrids, and islanded networks, where they enable grid stability and resilience while supporting broader goals of decarbonization and modernization \cite{GridForming2024}.  In this context, the present work proposes a novel IBR control approach that takes advantage of the flexibility of the converters to improve dynamic performance and grid stability by providing both frequency and voltage support through a coupled frequency-voltage controller.

\subsection{Literature review}

The theory concerning the synchronization, control, and stability of Synchronous Machines (SMs) has matured over more than a century.  This theory, in essence, relies on rotor speed and terminal voltage regulation achieved through adjusting the mechanical power and field voltage of the SMs.  These controllers are naturally decoupled by their time scales, with the latter being orders of magnitude faster.  Moreover, in conventional transmission systems ($X \gg R$) with a substantial share of SMs, the active and reactive power flows are strongly coupled with the frequency and magnitude of the voltage, respectively.  This further contributes, in conventional systems, to decouple active~power-frequency and reactive~power-voltage magnitude controllers.

On the other hand, the theory on synchronizing and regulating converters connected to the grid is relatively new and has been constantly updated along with the converter technology.  Synchronization in converters is not intrinsic, as in the case of SMs, but is forced by the estimation of the grid frequency, which is then utilized by the inner loops of the converter, which are typically related to current or voltage.  In recent years, the categorization of Grid-Forming (GFM) and Grid-Following (GFL) has been employed to understand the converter's synchronizing mechanism and its inner-loop structure.  Recent works have proposed a theoretical, although simplified, framework to classify these two structures \cite{TimGreen_Duality, osti_1813971}, while \cite{Sijia2022Unified} uses features from both structures to propose a unified mixed scheme.

Once synchronized, the main objectives of the control of the converter can be designed based on system requirements \cite{ancserv}.  Conventional structures for inverters outer loops typically include, among others, active and reactive power loops \cite{PQDecoupling}, voltage and frequency support through droop gains \cite{TOULAROUD2023Hier, Wei2023Suple, RAJAN2021PFC, OptimalDroop}, and virtual inertia \cite{Adaptive_Inertia_Control, Wei2023Suple}.
In terms of time scales, one can choose each control loop bandwidth depending on the control structure and its objectives.  For example, \cite{BandwidthSep} presents different effects on stability when choosing different speeds for the dynamics of the PLL, dc-link voltage, and the AC terminal voltage controls.  In the same vein, \cite{symcontrol_coupling} proposes a $\rm d$-axis feed-forward and $\rm q$-axis compensation method in the PLL to improve the asymmetry introduced to the frequency by the dc-link voltage controller.

Depending on the communication infrastructure, control objectives, and system dynamics, various control architectures can be considered. In \cite{FAZAL2023}, the authors provide a comparative analysis and review of popular techniques for centralized, decentralized, and hierarchical controllers. Although in this work we utilize multiple bus measurements, these are considered local since they are from adjacent nodes (e.g., taken at the secondary terminal of the IBR transformer). This characterizes the scheme as decentralized at the primary level, where it is used for regulating the output voltages and currents of the IBR.

In distribution systems ($X \approx R$) with low inertia and a significant share of converter-interfaced generation, there is a strong dynamic coupling between frequency and voltage variations and their controllers. Moreover, compared to conventional power systems, grids with a high share of IBRs exhibit similar time scales for frequency and voltage magnitude controls, and the coupling between active and reactive powers becomes more pronounced.  Some works have attempted to decouple this behavior by adjusting the voltage reference \cite{freqvoltage} and utilizing virtual impedance approaches \cite{PQDecoupling, VCFC_Impedance}.  Nevertheless, these approaches can be less effective in terms of overall dynamic response, as they often sacrifice voltage control to regulate frequency (or vice versa) or oversimplify the inherently coupled dynamics.  Alternatively, others exploit this natural coupling to enhance overall system stability \cite{bernal2024improving, Farrokhabadi2017, derfv, FastVoltageBoosters, curvatureCtrl, OptimalDroop}.  In this work, we adopt the latter strategy by proposing a control scheme that explicitly exploits the coupling between frequency and voltage controllers, as well as active and reactive powers, to improve overall grid stability.

Our starting point is the recently proposed concept of Complex Frequency (CF) (see the CF definition in Appendix \ref{sec:app_CF}). In simpler terms, CF can be defined as a variable or operator that captures the dynamics of a complex variable in a compact and consise form. Specifically in the context of power systems, the dynamics of voltages and currents, represented by their CF can be used to link their variations with the rate of change of complex power \cite{cmplx}. This relationship makes CF a powerful and versatile tool for analyzing power system dynamics.

The concept of CF has already been applied in various contexts, such as the classification of power converter control schemes \cite{moutevelis2023taxonomy}, the modeling of power system dynamics \cite{buttner2023}, the generalization of synchronization and droop control for grid-connected converters \cite{complexfrequencysync}, and the quantification of local dynamic performance \cite{bernal2024improving,derfv}.

Thus, in this work, we design a decentralized controller for IBRs that exploits the inherent and control-driven relationships between power and local variations in voltage phase and magnitude. This approach unifies the control of frequency and voltage regulation while improving overall dynamic performance, all within the CF framework.

\subsection{Contributions}

The main contributions of this work are as follows:
\begin{enumerate}
\item We propose a novel controller for IBRs based on the Complex Frequency approach.  Specifically, the controller objective is to maintain a constant CF for the voltage at the terminals of the point of connection of IBRs with the grid.
\item By enforcing a constant CF, the proposed controller achieves a unified and effective regulation of both the phase (angle) and magnitude of the voltage. The proposed approach differs from conventional schemes that decouple these dynamics, and exploits the potential arising from the inherent coupling between frequency and voltage magnitude variations.
\item We show that the proposed approach can be implemented in existing conventional controllers and we demonstrate its effectiveness in enhancing the overall dynamics of power systems through simulations of several scenarios and contingencies.  This improvement is observed in terms of its dynamic performance, thus, through local voltage frequency and magnitude regulation, as well as in the frequency of the Center of Inertia (CoI) of the system.
\end{enumerate}

\subsection{Organization}

The remainder of the paper is organized as follows.  Section \ref{sec:control} presents the proposed IBR $\eta$-control scheme derivation and implementation.  In Section \ref{sec:case}, two case studies are presented.  These are based on a modified version of the IEEE 9-bus system and a 1479-bus dynamic model of the Irish transmission grid.  Section \ref{sec:conclusion} draws conclusions and outlines future work.  Three Appendices complete the paper by providing theoretical background on the CF concept, the Park's vector and the link between current and voltage Park's vectors in ac branches.

\section{Proposed $\eta$-Control Scheme}
\label{sec:control}
In this section, we present the proposed control scheme, hereinafter denoted as \textit{$\eta$-control}.  The $\eta$-control utilizes the CF (see the definition in Appendix \ref{sec:app_CF}) as a dynamic reference to control voltage dynamics and reduce its variations.

We represent the voltage at the grid connection point using Park’s vectors in $\rm dq$ components.  Similarly, the current injected by the converter into the grid is also represented in $\rm dq$ components (see Appendices \ref{sec:dyn_park} and \ref{sec:dyn_curr}).  The main objective of $\eta$-control is to maintain a constant CF at the converter terminals.

The implementation of the $\eta$-control is detailed below, considering single and multiple adjacent nodes.  This section also discusses the implementation of a current limiter to manage device degradation and stability, enhancing the robustness of the control strategy.

\subsection{CF as a control reference}

The CF of a complex time-dependent quantity provides information about its rate of change in time and, hence, its dynamic behavior.  In the other way around, one can impose a dynamic for a given time-dependent quantity by setting a specific CF reference.  In particular, if we chose the objective of the controller to reduce variations of the voltage (assumed to be a Park's vector, that is, a complex time-dependent quantity), then we focus on setting a steady-state reference objective for the CF, given by:
\begin{equation}
  \overline{\eta}^{\mathrm{ref}} =
  \rho^{\mathrm{ref}}+ \jmath \, \omega^{\mathrm{ref}} = 0+\jmath \, \omega_o \, ,
  \label{eq.reference}
\end{equation}
where $\rho^{\mathrm{ref}}$ and $\omega^{\mathrm{ref}}$ are the real and imaginary part of the CF reference and $\omega_o$ refers to the system steady state frequency.  Setting the real part of the CF reference to null implies that the desired output for the voltage remains constant.
\subsection{Proposed control scheme}
\label{sec.control}
From the definition of the voltage as a complex vector, we can define its magnitude $v_h(t)$ and phase $\theta_h(t)$ in terms of $\rm dqo$ components, as follows:
\begin{equation}
  v_h(t) = \sqrt{v_{h,\mathrm{d}}(t)^2+v_{h,\mathrm{q}}(t)^2} \, ,
  \label{eq.rho_dq}
\end{equation}
and
\begin{equation}
  \theta_h(t) = \theta_{\mathrm{dq}}(t) +
  \arctan{\left(\frac{v_{h,\mathrm{q}}(t)}{v_{h,\mathrm{d}}(t)}\right)}\, ,
  \label{eq.omega_dq}
\end{equation}
where $\theta_{\mathrm{dq}}(t)$ is the phase angle of the $\rm dqo$-transform.  In the remainder of the paper, variable time dependence is dropped to improve legibility.  Note that all variables are assumed to be time-dependent unless stated otherwise.

Replacing (\ref{eq.rho_dq}) and (\ref{eq.omega_dq}) into the CF definition (refer to (\ref{eq.cmplx_freq}) in Appendix \ref{sec:app_CF}), we obtain the Park derivative operator $\overline{\mathrm{p}}$, as follows:
\allowdisplaybreaks
\begin{align}
  \nonumber
    &\overline{\eta}_h\overline{v}_h = \left(\frac{\dot{v}_h}{v_h}+j\dot{\theta}_h\right)\overline{v}_h \\
  \nonumber
    &=\left(\frac{\frac{d}{dt} \sqrt{v_{h,\mathrm{d}}^2+v_{h,\mathrm{q}}^2} }{\sqrt{v_{h,\mathrm{d}}^2+v_{h,\mathrm{q}}^2}} + \jmath \frac{d}{dt} \left[ \theta_{\mathrm{dq}} +
  \arctan{\left(\frac{v_{h,\mathrm{q}}}{v_{h,\mathrm{d}}}\right)}\right] \right) \overline{v}_h\\
  \nonumber
    &=\left(\frac{v_{h,\mathrm{d}}\dot{v}_{h,\mathrm{d}}+v_{h,\mathrm{q}} \dot{v}_{h,\mathrm{q}}}{v_{h,\mathrm{d}}^2+v_{h,\mathrm{q}}^2} + \jmath \frac{v_{h,\mathrm{d}}\dot{v}_{h,\mathrm{q}}-v_{h,\mathrm{d}}\dot{v}_{h,\mathrm{d}}}{v_{h,\mathrm{q}}^2+v_{h,\mathrm{q}}^2} \right) \overline{v}_h\\
  \nonumber
    &\ \ \ \ +\jmath \, \dot{\theta}_{\mathrm{dq}}\overline{v}_h \\
  \nonumber
    &= \dot{v}_{h,\mathrm{d}}+\jmath \, \dot{v}_{h,\mathrm{q}} + \jmath \, \omega_{\mathrm{dq}}\overline{v}_h\\
  \nonumber
    &= \left(\frac{d}{dt} + \jmath \, \omega_{\mathrm{dq}}\right) \overline{v}_h\\
    \label{eq.park2cf}
    &=\overline{\mathrm{p}}\ \overline{v}_h \, .
\end{align}

By using the equivalence of (\ref{eq.park2cf}) for the voltages in the
current flow dynamic equation (see (\ref{eq.diYv}) in Appendix
\ref{sec:dyn_curr}), and rearranging to express the derivative of the
current in $\rm dq$ components, we obtain:
\begin{equation}
  \frac{d}{dt}{\overline{\imath}_h} =
  \overline{Y}_{hk} ( \overline{v}_h  \overline{\eta}_h -
  \overline{v}_k   \overline{\eta}_k) -
  \jmath \, \omega_{\mathrm{dq}}  {\overline{\imath}_h }\, .
  \label{eq.di_complex}
\end{equation}

Equation (\ref{eq.di_complex}) is general and model-agnostic to the
device connected at bus $h$ or $k$.  This implies that understanding
the dynamics of the current does not require knowledge of the device's
model, as long as the voltage and its CF at each node are properly
defined.

By imposing a dynamic for the current reference given by
(\ref{eq.di_complex}) of a device connected to bus $h$, we can define
a control objective to maintain a constant CF, as defined in
\eqref{eq.reference}, at its terminals.  Similarly, achieving a
constant CF for the remote bus $k$ is also possible.  In the first
scenario, a remote measurement for the voltage at bus $k$ and its
corresponding CF is required, along with the voltage measurement for
the local bus $h$.  Accordingly, in the second scenario, the
estimation of the CF for the local bus would be required instead.

Equation (\ref{eq.di_complex}) accounts for the current derivatives of
a single device connected in antenna to the grid through, i.e., with a
single line connecting buses $h$ and $k$.  A generalization of
(\ref{eq.di_complex}) that considers a device connected to the grid
through multiple lines is as follows:
\begin{equation}
  \frac{d}{dt}{\overline{\imath}_h} = \sum_{k \in
    \mathcal{K}}{\overline{Y}_{hk} ( \overline{v}_h \overline{\eta}_h-
    \overline{v}_k \overline{\eta}_k) -\jmath \, \omega_{\mathrm{dq}}
    \overline{\imath}_{hk}} \, ,
  \label{eq.di_complex_mul}
\end{equation}
where $\mathcal{K}$ denotes the set of buses adjacent to bus $h$.
Note that (\ref{eq.di_complex_mul}) can also include shunt elements as
these can be modelled as admittances with the ground reference as an
adjacent node.

For the case of multiple adjacent bus bars, one can establish a
constant CF as a control objective for any specific bus bar, whether
local or adjacent.  This requires measuring all voltages and other CFs
aside from the one set as constant. 

The Park reference frequency, denoted as $\omega_{\mathrm{dq}}$, is also set to
enforce a constant CF in the current dynamics.  Thus, we set the park
reference to be equal to the imaginary part of the CF reference,
denoted as $\omega_o$.

By implementing the constant reference provided in
(\ref{eq.reference}) for the CF at bus $h$, we can derive the
generalization of the required current dynamics for a device connected
at bus $h$.  This current is then imposed as the reference input for
the inner current control, resulting in the final expression for the
proposed $\eta$-control, as follows:
\begin{equation}
    T_{\eta}\frac{d}{dt}\overline{\imath}_{\eta}^{\mathrm{ref}}=
    \sum_{k \in \mathcal{K}}{ \overline{Y}_{hk} ( \overline{v}_h
      \overline{\eta}^{\mathrm{ref}}- \overline{v}_k
      \overline{\eta}_k) -\jmath \, \overline{\omega}^{\mathrm{ref}}
      \overline{\imath}_{hk}},
    \label{eq.iref_control}
\end{equation}
where $T_{\eta}$ is included as a time constant on the left side to
adjust the speed of the current dynamic, and all values not designated
as a reference (with the super-index ``ref'') can be estimated based on
available measurements.

The admittance $\overline{Y}_{hk}$ is assumed to be a constant and a known
complex parameter. Nevertheless, by implementing the measured voltage
at each node and incorporating the current measurement of each branch
$\overline{\imath}_{hk}$, it is possible to estimate the admittance.
This allows us to modify expression (\ref{eq.iref_control}) as
follows:
\begin{equation}
    T_{\eta}\frac{d}{dt}\overline{\imath}_{\eta}^{\mathrm{ref}}= \sum_{k \in \mathcal{K}}{\left(\frac{\overline{v}_h \overline{\eta}^{\mathrm{ref}}- \overline{v}_k  \overline{\eta}_k}{\overline{v}_h-\overline{v}_k} -\jmath \, \overline{\omega}^{\mathrm{ref}}  \right)\overline{\imath}_{hk}}.
    \label{eq.iref_control_estY}
\end{equation}
%

\subsection{Implementation of $\eta$-control for IBRs}
\begin{figure}[htb]
  \centering
  \includegraphics[width = \linewidth]{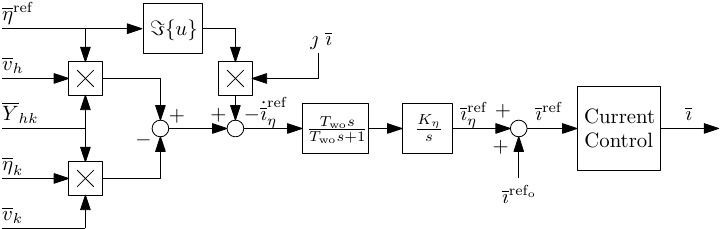}
  \caption{Block diagram of the proposed $\eta$ control.}
  \label{fig.block}
\end{figure}

Figure \ref{fig.block} shows the proposed control scheme.  The block diagram represents one possible implementation of \eqref{eq.iref_control} in the case of considering only one adjacent bus bar.  This scheme introduces three key modifications to enhance the controller's flexibility, as follows.
\begin{enumerate}
\item A gain $K_{\eta}$ is added to the integrator that calculates the variable $\overline{\imath}_{\eta}^{\mathrm{ref}}$ to adjust the speed of the control. In practice, this gain represents the inverse of the time constant $T_{\eta}$.
\item A wash-out filter is included before the integrator control to ensure support only during transient conditions.  This design choice enables parallel multi-instance operation, as it does not require perfect tracking for either frequency or voltage. 
\item An extra input $\overline{\imath}^{\mathrm{ref_o}}$ is included which can be associated with the initial condition for the current and a possible slower controller such as active and reactive power, voltage and frequency, power factor or a combination of the above, among other possible options.
\end{enumerate}

\begin{figure}[htb]
  \centering
  \includegraphics[width =0.725\linewidth]{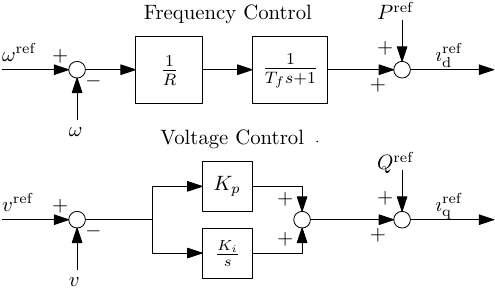}
  \caption{Block diagram of the conventional frequency and voltage controllers.}
  \label{fig.cc}
\end{figure}

For comparison, Fig.~\ref{fig.cc} shows the block diagram of a conventional frequency and voltage control loops.  The former consists of a droop gain $R$ and a first-order filter with a time constant $T_f$.  For the voltage control, a PI controller with gains $K_p$ and $K_i$ is utilised.  Finally, ${\imath}^{\mathrm{ref}}_{\mathrm{d}}$ and ${\imath}^{\mathrm{ref}}_{\mathrm{q}}$ denote the current references for the conventional inner-loop converter control, which may also be interpreted as inputs for the $\eta$-control, represented by the complex signal $\overline{\imath}^{\mathrm{ref_o}}$.

The $\eta$-control aims at minimizing complex frequency variations by maintaining a complex set point of $\eta^{\mathrm{ref}}=0+\jmath \omega_o$ pu/s.  For fast transients, both voltage magnitude and frequency try to remain constant by properly adjusting the complex current.  On the other hand, the conventional control independently adjusts the real and imaginary components of the current to control voltage frequency and magnitude via active and reactive power set points, respectively.  In conventional controllers the inherent interdependency between voltage magnitude and frequency, coupled with active and reactive power, makes this control inefficient and inexact during transients.

We consider a simplified current control for the IBR model and an ideal PLL synchronization to the grid, which regulates the $\rm d$ and $\rm q$ axis components of the current $\overline{\imath}=\imath_{\mathrm{d}}+\jmath \, \imath_{\mathrm{q}}$. For simplicity, we represent it as a first-order delay in both components, as follows:
\begin{equation}
  \begin{aligned}
    T_d\dot{\imath}_{\mathrm{d}} &= \imath_{\mathrm{d}}^{\mathrm{ref}}-\imath_{\mathrm{d}} \, , \\
    T_q\dot{\imath}_{\mathrm{q}} &= \imath_{\mathrm{q}}^{\mathrm{ref}}-\imath_{\mathrm{q}} \, .
  \end{aligned}
\end{equation}

\subsection{Current Limiter Implementation}
\label{sec.currlim}

Limiters are required for effectively managing device degradation and
stability control.  Unlike synchronous machines, converters typically
tolerate transient currents close to their nominal values.  These
limits are mainly determined by the current allowed through the
semiconductors \cite{ConverterTech}.

For stability purposes, proper current limiters have to be
incorporated into the controllers to prevent wind-ups within the
integrators and avoid inappropriate references for direct and
quadrature quotas.
Although there are many schemes for current limiters, such as those
based on virtual impedance through voltage references or direct
restrictions in the current \cite{GF_converters}, we have opted for a
circle limiter for this specific case.  This choice is motivated
because of its feature to maintain the angle unchanged
\cite{circlelim}.

\begin{figure}[htb]
  \centering
  \includegraphics[width = 88mm]{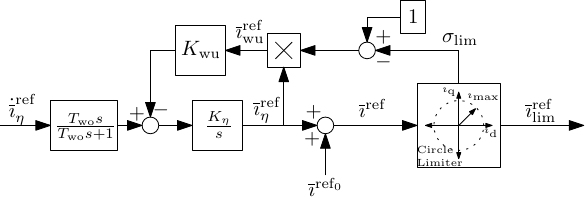}
  \caption{Block diagram of the current limiter.}
  \label{fig.lim_block}
\end{figure}

Figure \ref{fig.lim_block} illustrates the block diagram of a circular
current limiter designed for $\eta$-Control.  The expression for the
circular current limiter is given by:
\begin{equation}
  \overline{\imath}^{\mathrm{ref}}_{\mathrm{lim}} =
  \overline{\imath}^{\mathrm{ref}} \sigma_{\mathrm{lim}} \, ,
\end{equation}
where
\begin{equation}
  \begin{aligned}
    \sigma_{\mathrm{lim}} &= \min \left\{ 1,\frac{\imath_{\max}}{||\overline{\imath}^{\mathrm{ref}}||} \right\} \, , \\
    ||\overline{\imath}^{\mathrm{ref}}|| &= \sqrt{{\imath^{\mathrm{ref}}_\mathrm{d}}^2+{\imath^{\mathrm{ref}}_\mathrm{q}}^2} \, ,
  \end{aligned}
\end{equation}
and $\imath_{\max}$ is the maximum allowable converter current
magnitude.  $\sigma_{\mathrm{lim}}$ is a real-valued coefficient,
which is decreased only when the circular current limiter is
triggered, decreasing $\overline{\imath}^{\mathrm{ref}}$ magnitude
accordingly while its angle is kept unchanged.

Anti-windup limiters play a crucial role in preventing stability and
numerical issues.  To mitigate these problems, in some cases, it is
essential to include back calculation \cite{awulim}.  In this proposed
limiter, $K_{\mathrm{wu}}$ is the back calculation gain.  This gain
must be adjusted coherently with the speed of the main controller and
the dynamics of the controlled variable.
Back calculation loop is adjusted to reduce the magnitude of the
current $\overline{\imath}_{\eta}^{\mathrm{ref}}$ without changing the
angle, similar to what the circle limiter does.

\begin{figure}[htb]
  \centering
  \includegraphics[width = 0.50\linewidth]{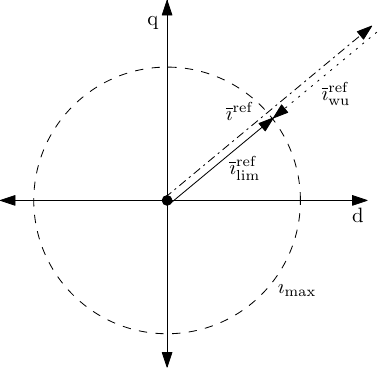}
  \caption{Vector diagram of the current limiter.}
  \label{fig.lim_vector}
\end{figure}

Figure \ref{fig.lim_vector} shows a vector plot in a $\rm dq$
reference frame representing the key variables of the circle
limiter.  In this scheme, it can be observed that the magnitude of
$\overline{\imath}_{\mathrm{lim}}^{\mathrm{ref}}$ is confined within a
circle of radius $\imath_{\mathrm{max}}$, regardless of the magnitude
of $\overline{\imath}^{\mathrm{ref}}$.  The variable
$\overline{\imath}_{\mathrm{wu}}^{\mathrm{ref}}$ denotes the current
used for back-calculation.

\section{Case Study}
\label{sec:case}

This section presents simulation results based on a modified version of the WSCC 9-bus test system \cite{Sauer_Book}, which has been adjusted for testing the dynamic performance of a single IBR; and the 1479-bus model of Irish transmission system, which is utilised to evaluate multiple IBR instances.  In both systems, the SMs are equipped with automatic voltage regulators and turbine governors.

The IBRs are modeled employing the proposed $\eta$-control, according to the details provided in Section \ref{sec.control}.  Additionally, a voltage and frequency outer loop is considered.  The parameters governing the current, frequency, and voltage controllers of the IBRs are outlined in Table \ref{tab.case_params_Der}. These parameters are adjusted through trial and error to ensure effective overall dynamic performance, and they remain unchanged for all cases unless explicitly stated otherwise. For voltage control, a simple PI controller is employed, while for frequency control, a combination of droop gain and a subsequent first-order filter is utilized.  Note that, although the fundamentals of the proposed control in this work can be applied to both GFM and GFL synchronization, in this case study, we consider exclusively GFL schemes for IBRs as these are the ones currently implemented in the Irish transmission system.

For each IBR, only one adjacent busbar is considered, specifically the one located at the grid side terminals of its transformer.  We thus assume that only two measurements are required for the proposed controller and that, unless state otherwise, measurements are not affected by delay.

When comparing with standard controllers, a control scheme which consists of an inner current control loop and two outer loops for frequency and voltage regulation is considered (refer to Figure \ref{fig.cc}). 

To evaluate dynamic performance, the index $\mu_r$ is used to quantify variations in voltage magnitude and phase at bus $r$ (refer to Appendix \ref{sec:mu} for its definition). Specifically, to assess the overall system dynamics, we define the index $\mu$ as:
\begin{align}
\mu = \sum_{r \in \mathcal{R}}{\mu_r},
\end{align}
where $\mathcal{R}$ represents the set of all buses in the grid. This index provides a scalar measure of the total voltage variation across the entire system.

\begin{table}[H]
\caption{IBR Controller Parameters}
\centering
\begin{tabular}{ll}
\hline
Controller & Parameters \\
\hline
Current & $T_d = 0.001$ s, $T_q = 0.001$ s \\
Frequency & $R = 0.06$, $T_f = 1.2$ s \\
Voltage & $K_i = 5$, $K_p = 10$ \\
\hline
\end{tabular}
\label{tab.case_params_Der}
\end{table}

\begin{figure}[htb]
  \centering
  \includegraphics[width =0.575\linewidth]{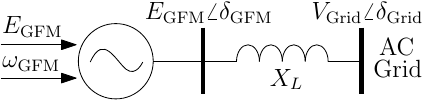}
  \caption{ GFM controllable voltage source behind a coupling reactance.}
  \label{fig.GFM_source}
\end{figure}

\begin{figure}[htb]
  \centering
  \includegraphics[width =0.625\linewidth]{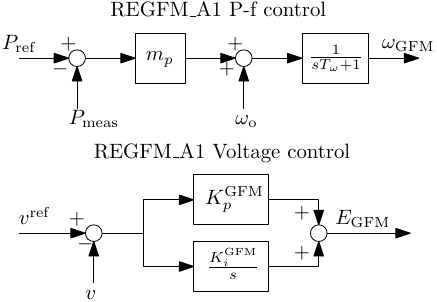}
  \caption{Block diagram of the internal voltage magnitude $E_{\mathrm{GFM}}$ and angular frequency $\omega_{\mathrm{GFM}}$ controllers for the REGFM\_A1 model.}
  \label{fig.GFM_control}
\end{figure}

In some scenarios, we also consider a GFM control scheme that provides virtual inertia.  With this aim, we utilize the REGFM\_A1 model described in \cite{REGFM_A1}.
The GFM inverter operates as a controllable voltage source behind a coupling reactance, as illustrated in Figure \ref{fig.GFM_source}.  The internal voltage magnitude $E_{\mathrm{GFM}}$ and angular frequency $\omega_{\mathrm{GFM}}$ are regulated by the controllers depicted in Fig.~\ref{fig.GFM_control}.  If the time constant $T_{\omega}$ is zero, then the P-f control behaves as a droop; otherwise, the P-F control resembles a synchronous machine, where the inertia is $M = T_{\omega}/m_p$, and the damping is $D = 1/m_p$.  The parameters used for the REGFM\_A1 model are outlined in Table \ref{tab.case_params_GFM}. 

\begin{table}[H]
\caption{REGFM\_A1 Controller Parameters}
\centering
\begin{tabular}{ll}
\hline
Controller & Parameters \\
\hline
P-f & $T_{\omega} = 20$ s, $mp = 0.001$ pu \\
Voltage & $K_i^{\mathrm{GFM}} = 1$, $K_p^{\mathrm{GFM}} = 5$ \\
Interface & $X_L = 0.15$ pu \\
\hline
\end{tabular}
\label{tab.case_params_GFM}
\end{table}

All simulation results presented in this section were obtained using
the simulation software tool Dome \cite{dome}.

\subsection{WSCC 9-bus System}

\begin{figure}[htb]
  \centering \includegraphics[width =
    82mm]{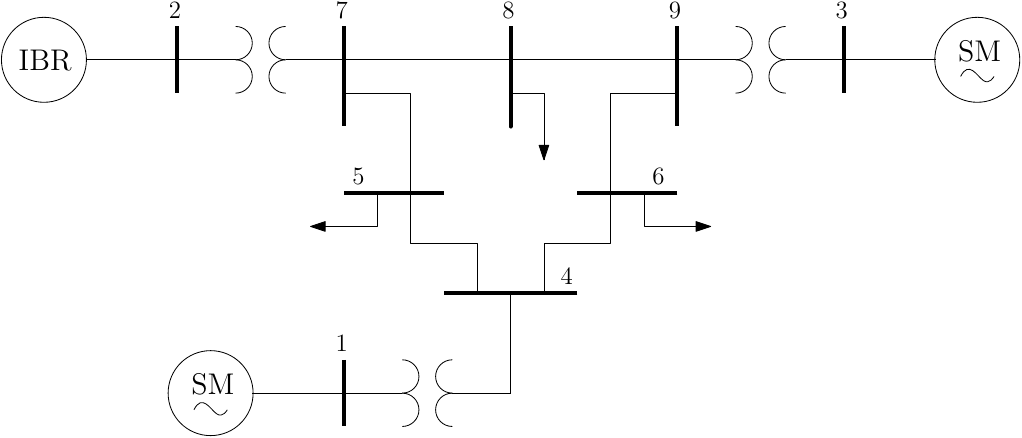}
  \caption{Modified WSCC 9-bus system.}
  \label{fig.case_wscc}
\end{figure}

We utilise the modified version of the WSCC 9-bus system, where the SM originally located at bus 2 has been replaced with an IBR.  The two remainder SMs are represented with a 4th order (two-axis) model.  The single-line diagram depicting the modified system is presented in Fig.~\ref{fig.case_wscc}.  The configuration for the $\eta$-control gain is determined to be $K_{\eta}=1$ and a high wash-out filter of $T_{\mathrm{wo}}=50$ seconds is considered, these values are  consistently applied across all scenarios discussed in this subsection.

In the remainder of this section, we discuss the performance of the
proposed control scheme following a load outage and a three-phase
fault. Additionally, the impact of limits, delays, estimation, R/X ratio and EMT dynamics is
examined. Note that all quantities are expressed in per unit.
Specifically, active and reactive power are referenced to 100 MVA,
while currents are based on the same power but a voltage base of 18
kV, which corresponds to the IBR connection voltage level.

\subsubsection{Power Unbalance}

\begin{figure}[htb]
  \centering
  \includegraphics[width = 88mm]{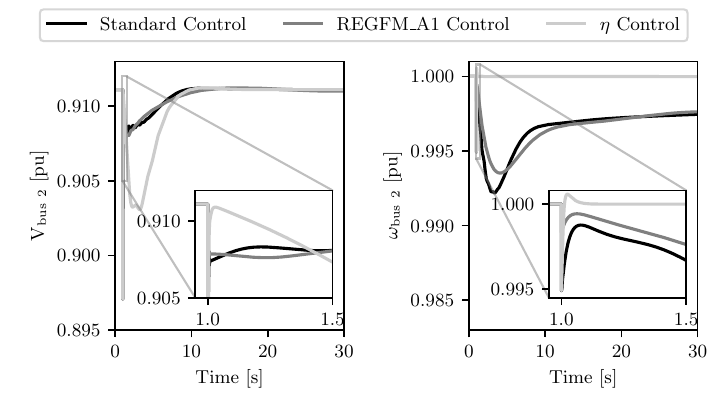}
  \caption{WSCC 9-bus system --- Positive load step at bus 5 --- Voltage (left panel) and frequency (right panel) at bus 2 for different control setups of the IBR.}
  \label{fig.load_v_omega_wscc}
\end{figure}

We consider an increase in the load located at bus 5.  A positive step change equivalent to a 16\% of the total active power load is applied at the first second of simulation. Figure \ref{fig.load_v_omega_wscc} depicts the voltage magnitude and frequency measured at bus 2, the bus to which the IBR is connected.  A comparison of the performance between the Standard and $\eta$-control is conducted.  The dynamic behavior of the system over 30 s of simulation is displayed.  

The voltage dynamics result from a combination of conventional PI voltage control and $\eta$-control. The $\eta$-control primarily focuses on minimizing transient voltage variations at high frequencies, whereas the PI control, being slower, responds to the voltage error within the limits imposed by the $\eta$-control.

An AGC is included in the system model to recover a frequency of 1 pu in steady state.  The AGC signals are exclusively linked to the SMs governors.  However, a potential challenge arises for $\eta$-control if the IBR significantly increases its energy generation to stabilize the frequency.  In this case, the AGC will not detect any frequency change unless the IBR decreases its power injection.  The amount of power supplied and the duration for which the IBR sustains it directly depends on its available reserve and technical limitations.  This power supply coordination should be properly addressed within the active power outer-loop dynamics, which can be either linked with the AGC or manually adjusted to synchronize the reserve usage and the ramp-up/down of available power resources in the rest of the system.

\begin{figure}[htb]
  \centering
  \includegraphics[width = 88mm]{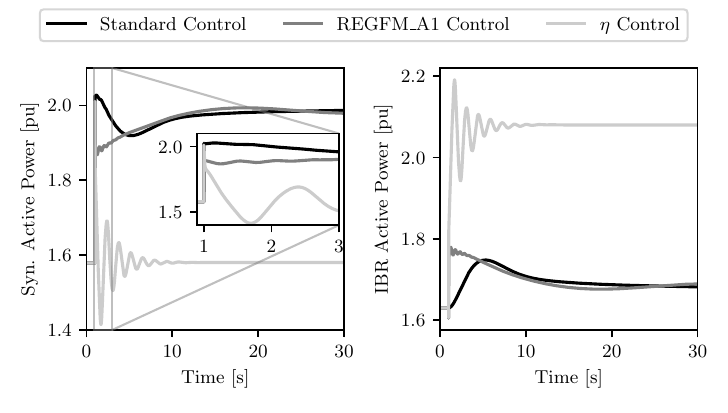}
  \caption{WSCC 9-bus system --- Positive load step at bus 5 ---
    Active power provided by the SMs (left panel) and active power
    provided by the IBR (right panel) to the system for different
    control setups of the IBR.}
  \label{fig.load_P_wscc}
\end{figure}

Figure \ref{fig.load_P_wscc} illustrates the active power supplied by the IBR and the combined output of the two remaining SMs in response to the load step.  In both cases, the SM presents an inertial response, which can be observed almost as an instantaneous step in the active power.  The effect of the IBR controllers manifest after the initial response of the SMs.

In the case of conventional control, the SM predominantly relies on inertial response, causing a deceleration before the primary frequency control takes effect through the governor.  This control mechanism stops the frequency from decreasing, and afterwards, operating on a much slower time scale, the AGC provides secondary frequency control aimed to recover inertial energy and gradually restore the frequency to its nominal value. 

In the case of the virtual inertia approach, we observe a similar behavior. However, the initial RoCoF is reduced, and the frequency nadir is improved due to the initial power response of the IBR to the frequency variation.

With the $\eta$-control, almost simultaneously with the inertial response, the IBR compensates for the load step by providing the energy required to reduce voltage variations.  Consequently, the SM avoids using its kinetic energy, thereby maintaining the frequency consistently close to its nominal value.

The dimensionless dynamic performance indices, $\mu$ (representing overall system performance) and $\mu_2$ (specific to local performance at bus 2), evaluated at 5 seconds, are presented in Table~\ref{tab:mu_load}. These indices are normalized relative to the performance under Standard control. The results demonstrate that the proposed controller achieves improved dynamic performance both locally and globally under load imbalance conditions.
\begin{table}[h!]
\centering
\caption{WSCC 9-bus system — Positive load step at bus 5 — $\mu@5sec$.}
\begin{tabular}{cccc}
Index & Standard & REGFM\_A1 & $\eta$ Control \\
\hline
$\mu_2@5sec$ & 1 & 0.806 & 0.020 \\
$\mu@5sec$ & 1 & 0.806 & 0.022  
\end{tabular}
\label{tab:mu_load}
\end{table}

The significant reduction in the index $\mu$, to less than 3\% of that of the standard control, is primarily due to the fast frequency regulation provided by the $\eta$ control. This capability allows the frequency to remain consistently close to its steady-state value, thereby minimizing the performance index compared to other simulated controllers.

\subsubsection{Three-phase Fault}

\begin{figure}[htb]
  \centering
  \includegraphics[width = 88mm]{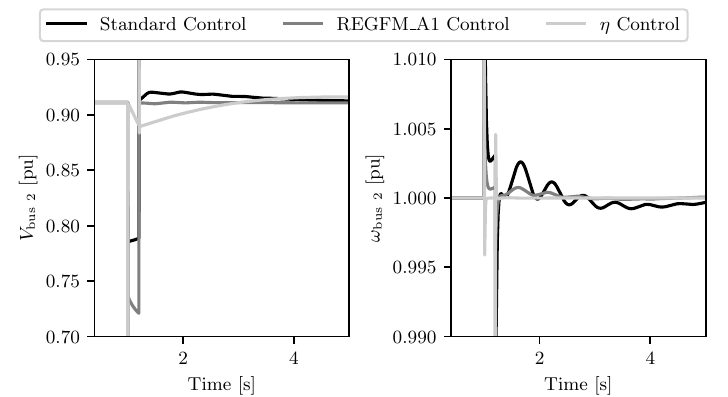}
  \caption{WSCC 9-bus system --- Three-phase fault at bus 7 ---
    Voltage (left panel) and frequency (right panel) at bus 2 for
    different control setups of the IBR.}
  \label{fig.fault_wscc}
\end{figure}

\begin{figure}[htb]
  \centering
  \includegraphics[width = 88mm]{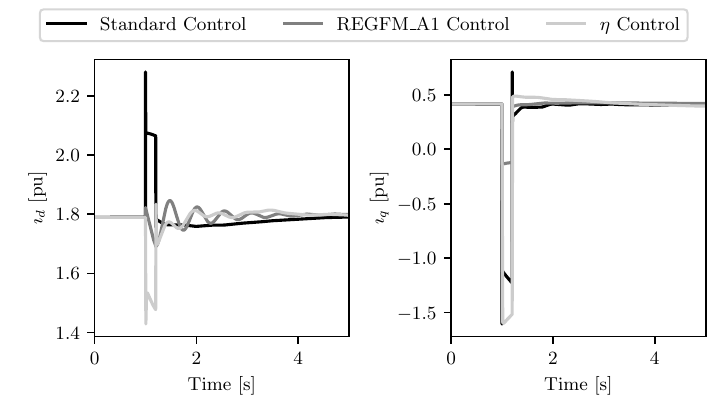}
  \caption{WSCC 9-bus system --- Three-phase fault at bus 7 --- Direct     current (left panel) and quadrature current (right panel) injected     at bus 7 for different control setups of the IBR.}
  \label{fig.fault_wscc_I}
\end{figure}

In this section, the proposed controller is compared with a conventional controller by analyzing the response under a three-phase fault located close to the IBR, at bus 7.  The fault has an impedance of $0.03+j0.3$ pu, which implies a voltage drop up to 0.7 pu at bus 7, the fault is applied for 200 ms.  Figure \ref{fig.fault_wscc} depicts the voltage magnitude and frequency measured at bus 2, while Figure \ref{fig.fault_wscc_I} depicts the direct and quadrature currents injected by the IBR.  In this case, only 5 s are shown to focus on the dynamic during the fault and a few seconds after the fault transient.

As the fault occurs, the controllers come into action.  It is interesting to observe that the $\eta$-control reduces its direct current, while the conventional and virtual inertia controllers increase it.  In terms of the quadrature current, the controllers increase their injections (considered negative by convention).  During the fault, the $\eta$-control significantly reduces voltage and frequency variations better than the other schemes.  The difference in behavior is mainly due to the fact that the complex controller treats both voltage magnitude and angle as a single complex variable for control, whereas the others neglect this interdependency, resulting in an incomplete or inefficient response.

The dimensionless dynamic performance indices, $\mu$ (representing overall system performance) and $\mu_2$ (specific to local performance at bus 2), evaluated at 5 seconds, are presented in Table~\ref{tab:mu_fault}.  The indices are normalized relative to the performance under Standard control.  These results highlight that the proposed controller effectively enhances dynamic performance, both at the system level and locally, in the presence of the fault.

\begin{table}[h!]
\centering
\caption{WSCC 9-bus system - Three-phase fault at bus 7 - $\mu@5sec$.}
\begin{tabular}{cccc}
Index & Standard & REGFM\_A1 & $\eta$ Control \\
\hline
$\mu_2@5sec$ & 1 & 0.416 & 0.19 \\
$\mu@5sec$ & 1 & 0.441 & 0.178 
\end{tabular}
\label{tab:mu_fault}
\end{table}

In this case, the performance index for the $\eta$-control is reduced to 18\% of that of the standard control. This significant improvement is primarily due to the inherent reduction in the oscillations following the fault, both in frequency and in voltage magnitude.

\subsubsection{Impact of current limiters}

This section discusses the impact of current limiters.  For the
proposed control, the limiters are included as described in Section
\ref{sec.currlim}.  In this particular case, the gain values for the
circle current back-calculation are set as $K_{\mathrm{wu}}=100$,
ensuring they are sufficiently fast to mitigate any dynamic issues
related with wind-up.

\begin{figure}[htb]
  \centering
  \includegraphics[width = 88mm]{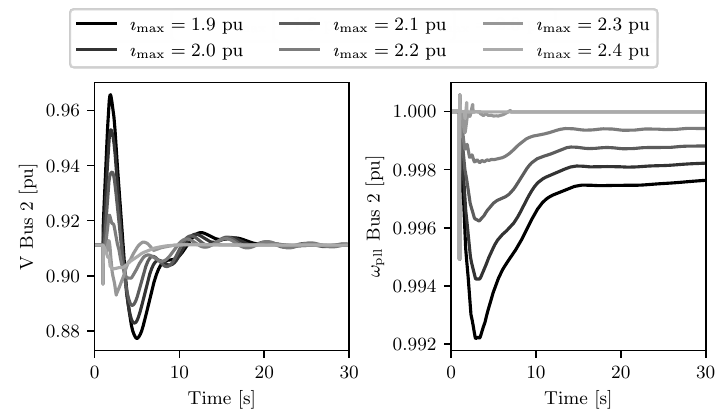}
  \caption{WSCC 9-bus system --- Positive load step at bus 5 --- Voltage (left panel) and frequency (right panel) at bus 2 for different current limits within the control.}
  \label{fig.lim_wscc}
\end{figure}

\begin{figure}[htb]
  \centering
  \includegraphics[width = 88mm]{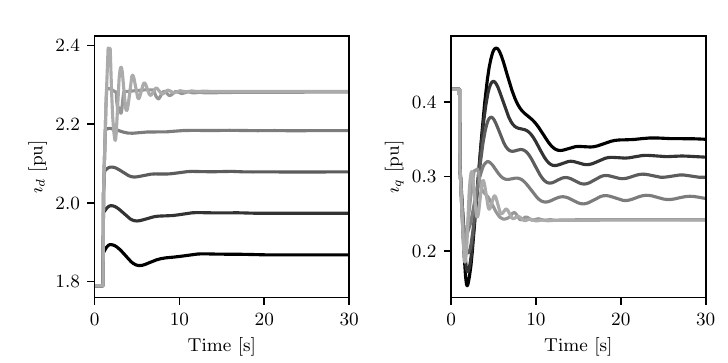}
  \caption{WSCC 9-bus system --- Positive load step at bus 5 --- Direct current (left panel) and quadrature current (right panel) at bus 2 for different current limits within the control.}
  \label{fig.lim_wscc_I}
\end{figure}

Figure \ref{fig.lim_wscc} presents the voltage magnitude and frequency at bus 2 for a positive load step equivalent to a 16\% of the active power load of the system.  Different current limits are considered, ranging from 1.9 to 2.4 pu, with steps of 0.1 pu.  Figure \ref{fig.lim_wscc_I} illustrates the direct and quadrature currents of the IBR.

The $\eta$-control attempts to control the CF of the voltage depending on the direct and quadrature current available from the circle limiter.  However, when reaching a limit, the interrelation between active and reactive power (which are strongly related with $\imath_{\mathrm{d}}$ and $\imath_{\mathrm{q}}$) with the frequency and magnitude of the voltage, involves a continuous control trade-off for these last two variables. For example, if the current approaches its limit and the controller requires a significant quota of direct current compared to quadrature current, the latter is reduced to maintain a vector current control with a defined direction. This phenomenon can be observed in the case where $\imath_{\text{max}}=1.9$ pu. Initially, the quadrature current is reduced to prioritize direct current, and as an initial compromise to sustain the frequency of the voltage, it affects its magnitude, enforcing a subsequent increase in priority for quadrature current.

If we relate these limits to a power reserve by considering them as the maximum output power that the IBR can deliver, then the 1.9 pu current limit would correspond to a power reserve of approximately 14\%, and a 2.4 pu limit would correspond to approximately 32\%.  While there is a trade-off between effectiveness and available power reserve, the controller consistently attempts to minimize variations with the available resources.

\subsubsection{Impact of delay in the remote measurement}

In this subsection, we study the effects of delays by including a pure delay in the measurement of the remote voltage.  This delay can become of significance when the remote measurement is located at a considerable distance, such as along a long transmission line, or when the communication system has other limitations.  In the case of the modified WSCC 9-bus system (refer to Figure \ref{fig.case_wscc}), for the IBR connected at bus 2, bus 2 would refer to the local voltage $\overline{v}_h$ and bus 7 to the remote voltage $\overline{v}_k$. Accordingly, the admittance $\overline{Y}_{hk}$ is related to the transformer that links buses 2 and 7.

\begin{figure}[htb]
  \centering
  \includegraphics[width = 88mm]{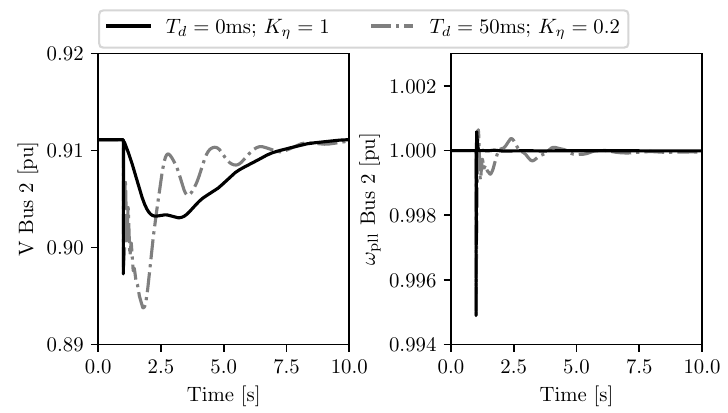}
  \caption{WSCC 9-bus system --- Positive load step at bus 5 --- Voltage magnitude (left panel) and frequency (right panel) at bus 2 for different remote measurement delays and control gains.}
  \label{fig.delay_wscc}
\end{figure}

Figure \ref{fig.delay_wscc} shows the voltage magnitude and frequency at bus 2 when a positive step equivalent to 16\% of the total load active power is applied in the load located at bus 5 at the first second of the simulation. Two cases are presented: (i) no time delays ($Td = 0$ ms; $K_{\eta} = 1$); and (ii) with a delay of 50 ms ($Td = 50$ ms; $K_{\eta}= 0.2$).  $\eta$-control gains must be adjusted to ensure that the control speed aligns accordingly with the delay and avoid stability and numerical issues.

As depicted in Figure \ref{fig.delay_wscc}, a significant delay negatively impacts the overall voltage magnitude and frequency response of the controller. Nevertheless, it's worth noting that this delay is often associated with long transmission lines (over 100 km). Thus, it becomes negligible when measurements are taken on both sides of a transformer, typically within distances under 1 km.
In case it is required, control settings can be fine-tuned to reduce the impact of this delay, by prioritizing local voltage control over $\eta$-control.

\subsubsection{Impact of CF estimation and noise}

The estimation of both the real and imaginary parts of the frequency is fundamental for a proper performance of the $\eta$-control. To estimate these variables we use a conventional synchronous reference frame Phase Locked Loop (PLL) for $\omega_h$ and another one for $\rho_h$.

\begin{figure}[htb]
  \centering
  \includegraphics[width = 0.85\linewidth]{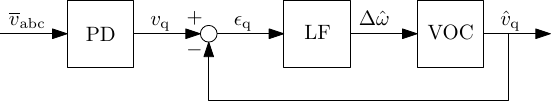}
  \caption{Scheme of a basic PLL.  PD is the phase-detector; LF is the loop filter; and VOC is the voltage-oscillator control.  The output of the LF can be utilised as an estimation of the frequency variations of the tracked voltage.}
  \label{fig.pll}
\end{figure}

Figure \ref{fig.pll} illustrates a basic scheme for a PLL, which is composed by a Phase Detector (PD), Loop Filter (LF) and a Voltage Oscillator Control (VOC) \cite{PLL}.  We use a standard synchronous-reference frame PLL model where the PD is modeled as a delay, the LF as a PI controller and a VOC is implemented as an integrator.  The parameters used for the estimation of $\omega_h$ are $K_{\mathrm{P}}^{\omega_h}=0.05$ and $K_{\mathrm{I}}^{\omega_h}=0.1$, while for the estimation of $\rho_h$ we use $K_{\mathrm{P}}^{\rho_h}=250$ and $K_{\mathrm{I}}^{\rho_h}=500$.

The dynamic behavior of the proposed controller is then evaluated in
the presence of varying loads.  These loads are characterized by
including additive noise to a constant load.  The noise is modeled
as a stochastic variable with a zero mean and a standard deviation of
5\% of the power of the constant load.
To study the impact of the load noise in the voltage estimation,
multiple simulations of over a 100 seconds are conducted, maintaining
a consistent noise seed, but increasing the noise speed in each
simulation.  The noise speed (or auto-correlation coefficient),
quantified in pu/s, refers to the rate at which the noise reaches a
steady state value of its standard deviation.
Finally, we compare the overall performance with the conventional
controller, which, although it needs a PLL for the frequency control,
it does not need to estimate $\rho_h$.

While power increments for stochastic loads are modeled up to 0.1 pu
in recent literature for the WSCC 9-bus system, with a granularity of
one sample per second in measurements \cite{PQStc}, our analysis
extends to load variations in each time simulation step. Here,
increments are modeled with a noise speed ranging from 0 to 20 pu/s.
This means that for a time step of 1 ms, a maximum change of 0.04 pu
is allowed.  Considering a standard deviation of 5\%, the total
increment per second is always constrained between $\pm$0.15 pu
($\pm$3 standard deviations).  This choice allows us to focus on
high-frequency variations within one second that may negatively impact
the estimation of the CF of the voltage while maintaining bounded load
increments.

\begin{figure}[htb]
  \centering
  \includegraphics[width=0.65\linewidth]{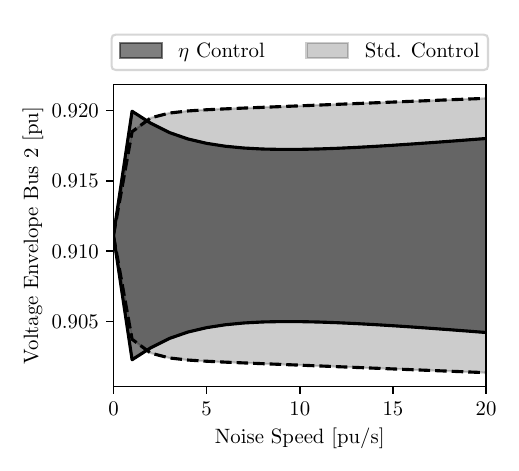} 
  \includegraphics[width=0.65\linewidth]{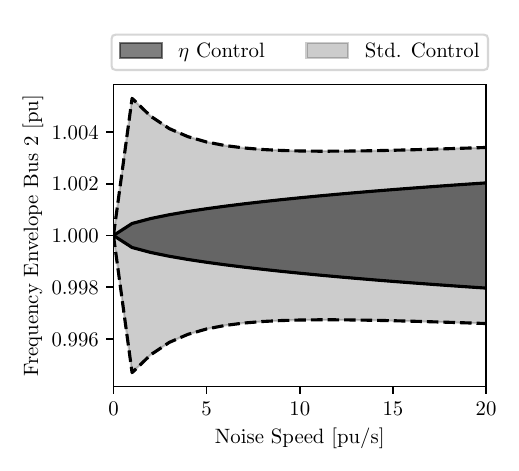}
  \caption{WSCC 9-bus system --- Voltage (left panel) and frequency
    (right panel) envelope at bus 2 represented as the mean $\pm$3
    std.~deviations for different control setups of the IBR as the
    noise of the stochastic loads increases.}
  \label{fig.noise_v_wscc}
\end{figure}

Figure \ref{fig.noise_v_wscc} illustrates the voltage magnitude and
frequency envelope at bus 2, respectively, as the noise speed
increases.  The envelope is defined as the mean value $\pm$3 times the
standard deviation of the variable, providing a graphical display to
represent the range of variations for all the simulations.

For the voltage magnitude, it can be observed that the conventional control exhibits higher variations compared to the $\eta$-control. However, this behavior is primarily attributed to the voltage derivative estimation of the $\eta$-control through the PLL. As part of its design, the PLL introduces a non-zero error during transients.
Regarding the frequency of the voltage, although variations increase overall with noise speed, those observed in the proposed controller consistently remain lower than those seen in the conventional control. In this scenario, both controllers rely on the PLL to estimate the angle and frequency of the voltage.

\subsubsection{Impact of R$/$X ratio}

\begin{figure}[htb]
  \centering
  \includegraphics[width = 88mm]{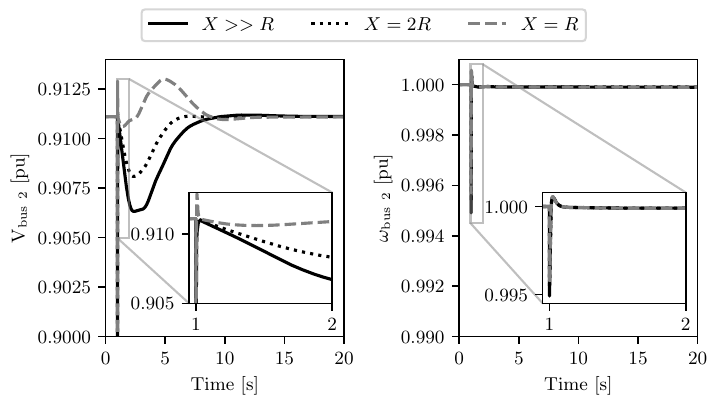}
  \caption{WSCC 9-bus system --- Positive load step at bus 5 --- Voltage magnitude (left panel) and frequency (right panel) at bus 2 for different R$/$X ratios.}
  \label{fig.rx_wscc}
\end{figure}

In this subsection, we simulate different $R/X$ ratios of the transmission lines of the modified WSCC 9-bus system. For each line, we maintain the magnitude of its series impedance while modifying the $R/X$ ratio. Three cases are evaluated: $X >> R$ (base), $X = 2R$, and $X = R$. Figure \ref{fig.rx_wscc} depicts the voltage magnitude and frequency at bus 2 for a load step at bus 5 for the three cases. It can be observed that while the frequency response remains almost unchanged, there is an improvement in the transient of the magnitude of the voltage.
For the case where $X = R$, the voltage stays close to the reference, then increases smoothly, and finally returns to the reference voltage. This is the result of a combined action of the $\eta$ control and the PI voltage control. The $\eta$ control affects both active and reactive power. Moreover, in the case of $X/R = 1$, active and reactive powers significantly impact voltage, making the combined response more effective.

\subsubsection{Impact of EMT Dynamics}

\begin{figure}[htb]
    \centering
    \includegraphics[width = 88mm]{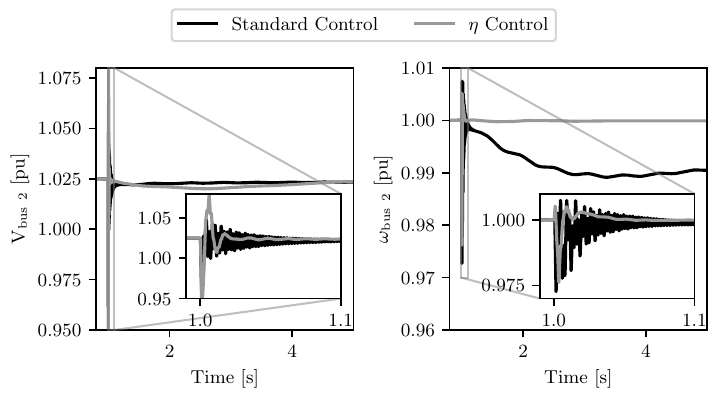}
    \caption{ Dynamic Phasors simulation for the WSCC 9-bus system --- Positive load step at bus 5 --- Voltage (left panel) and frequency (right panel) at bus 2 for different control setups of the IBR.}
    \label{fig:emt}
\end{figure}

In this section, we carry out simulations using a dynamic-phasor model of all devices.  The goal is to show that the proposed $\eta$-control scheme is robust and does not dynamically couple with fast dynamics.  For this scenario, we included the dynamics of the lines and transformers, PLL dynamics for synchronizing the IBR, and flux dynamics for the synchronous machine.  Overall the considered dynamic-phasor model is equivalent to a balanced EMT simulation with average models for the converter switching circuits.  The integration step is set equal to 50 $\mu$s to properly capture the fast dynamics of the system.  Figure \ref{fig:emt} illustrates a dynamic-phasor simulation for the WSCC 9-bus system, following a load increase step at bus 5.  This figure shows the voltage magnitude (on the left panel) and the frequency (on the right panel) at bus 2 for different control setups of the IBR.  Both controllers maintain synchronism, but the $\eta$-control dampens the faster dynamics more effectively than the standard scheme.  It is important to note that this behavior is achieved by employing a standard PLL for synchronization, which has been tuned through trial and error and remains consistent for both standard and $\eta$ controllers.  Specifically, the PI parameters for the PLL are set as $K_p=0.2$ and $K_i=1$.

\subsection{Irish grid}

In this section, we use the all-island Irish transmission system described in \cite{IRISH} to evaluate the performance of the proposed control considering multiple IBRs.  The grid comprises 22 synchronous generators, contributing 38\% of the total power (1080MW), 169 wind plants with a 30\% share (860MW), and interconnectors accounting for 32\% (906MW).  Additionally, there are 246 loads, 1479 buses, 796 lines, and 1055 transformers modeled within the system.
Synchronous generators are modeled using Sauer and Pai's 6th order machine, while wind turbines based on DFIG technology are replaced with IBRs.  The remaining wind generators are considered as constant-speed induction generator with a 5th-order squirrel-cage induction generator model.  While simulations are based on realistic data, they do not represent any specific operational condition of the system.

\begin{figure}[htb]
  \centering
  \includegraphics[width = 88mm]{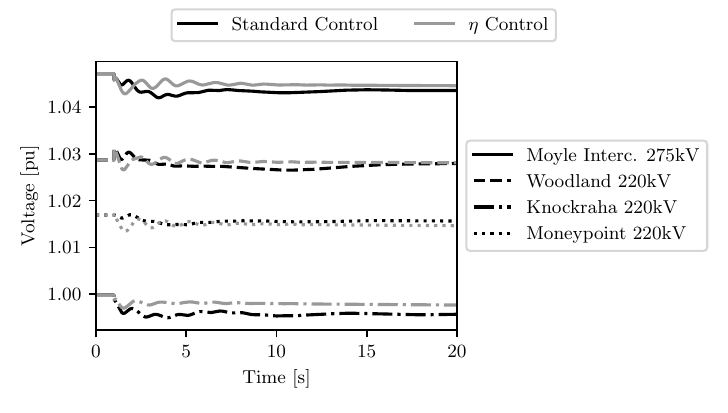}
  \caption{Irish system --- Voltage at representative buses after the loss of the East to West Interconnector 500 MW for different control setups of the IBR.}
  \label{fig.Ire_V}
\end{figure}

\begin{figure}[htb]
  \centering
  \includegraphics[width = 88mm]{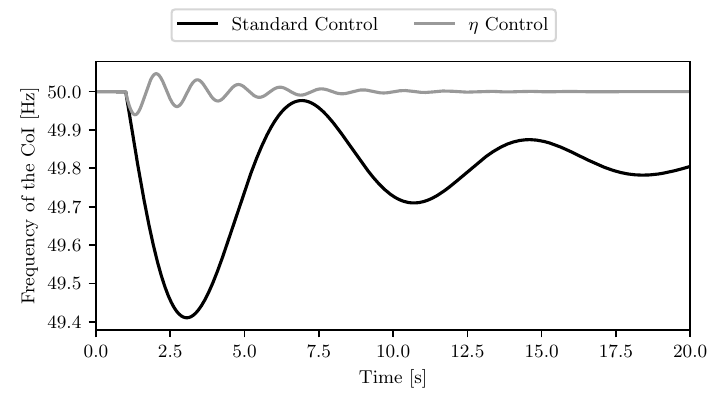}
  \caption{Irish system --- Frequency of the center of inertia after the loss of the East to West Interconnector 500 MW for different control setups of the IBR.}
  \label{fig.Ire_w}
\end{figure}

Figure \ref{fig.Ire_V} shows the voltage for various representative
substations of the Irish grid located in the north, west, east, and
south of the system, whereas Figure \ref{fig.Ire_w} shows the
frequency of the center of inertia for the Irish grid.  The
contingency consists in loss of the East to West interconnector, which
is importing its nominal active power of 500 MW.  This is one of the
most severe contingencies that can occur in the Irish transmission
system.  In this scenario, we assume that the IBRs have enough reserve
to effectively deliver the active and reactive power loss.  Finally,
$\eta$-control is tuned with the gain set to $K_{\eta}=10$.


Although both controllers meet the Irish grid technical requirements for frequency and voltage during transient system disturbances ([48, 52] Hz and [200, 245] kV for 220 kV rated busbars \cite{eirgrid2024gridcode}), an overall improvement in the magnitude of the voltage across the buses of the system is observed.  A significant difference in the frequency response of the Center of Inertia (CoI) is evident.  With $\eta$-control, the frequency of the CoI varies by less than 50 mHz, whereas with conventional control, it reaches nearly 600 mHz.

The system reaches its lowest frequency before 3 seconds under standard control, exhibiting an oscillating mode of approximately 0.11 Hz. In contrast, with $\eta$-control, the nadir is reached within 400 ms, and the system oscillates at approximately 0.7 Hz. While both modes can be classified as inter-area oscillation modes primarily influenced by the synchronous machines (SMs), the system dynamics are notably faster with $\eta$-control.

\section{Conclusions}
\label{sec:conclusion}
This paper proposes a novel $\eta$-control scheme to enhance the system stability and performance of IBRs.  The control utilizes the CF as a dynamic reference to maintain a constant CF at terminals, focusing on minimizing voltage variations.  The derivation covers single and multiple adjacent nodes, integrating a circular current limiter for practical implementation and stability enhancement.

Simulation of a variety of scenarios and contingencies is conducted within the WSCC 9-bus system and a 1479-bus dynamic model of the Irish transmission grid.  The $\eta$-control consistently outperforms the conventional control in response to power unbalance scenarios, such as load step changes, by allowing the IBRs to efficiently compensate for sudden load variations, maintaining the system frequency close to its nominal value, and minimizing dependence on inertial response.  Moreover, the dynamic performance index $\mu$ is reduced to less than 20\% of that of the standard control during critical contingencies in the WSCC 9-bus system. This result indicates the effectiveness of the $\eta$ control in minimizing voltage and frequency variations compared to conventional controllers.

Current limiters are considered to study their impact on the stability, demonstrating that the $\eta$-control adapts dynamically to varying current limits, optimizing resource usage based on the available power reserve.  The effects of delays in remote measurements is also analyzed, revealing that the $\eta$-control is sensitive to delays, but the control can be adjusted to mitigate their impact.  Finally, the analysis of stochastic load variations with noise, R/X ratio and EMT dynamics emphasizes the robustness of the $\eta$-control in minimizing voltage and frequency fluctuations compared to conventional schemes.

In the case of the Irish transmission system, the $\eta$-control demonstrates remarkable improvements in the frequency response and voltage stability, even after the loss of the largest infeed, i.e., the EWIC interconnector feeding 500 MW to the Irish grid.  The frequency of the CoI remains more stable, but exhibits faster dynamics with $\eta$-control.

Future work will focus on considering other IBRs configurations, such as GFM structures, as well as at testing the proposed controller in HIL simulations.

\vspace{-3mm}

\appendices

\section{Complex Frequency}
\label{sec:app_CF}
The complex frequency (CF) is a physical quantity that can also act as a derivative operator of any complex number with non-null magnitude \cite{cmplx}.  For example, considering a complex time-dependent quantity, say $\overline{u}(t)$, this can be written as:
\begin{equation}
  \overline{u}(t) =
  u(t) \exp(\jmath \, \alpha(t)) =
  \exp(\ln{u(t)}+\jmath \, \alpha(t)) \, ,
\end{equation}
where $u(t) \ne 0$ and $\alpha(t)$ are the magnitude and phase angle of $\overline{u}(t)$, respectively.  Assuming $\ln{u(t)}$ and $\alpha(t)$ are smooth functions of time, the time derivative of $\overline{u}(t)$ gives:
\begin{equation}
  \begin{aligned}
    \dot{\overline{u}}(t)
    &=\left (\frac{d}{dt}\ln{u(t)} +
    \jmath \, \dot{\alpha(t)} \right )\exp(\ln{u(t)}+\jmath \, \alpha(t))\\
    &=\left(\frac{\dot{v}_h(t)}{v_h(t)}+\jmath \, \dot{\alpha(t)}\right)\overline{u}(t) \\
    &=\left(\rho(t)+\jmath \, \omega(t) \right)\overline{u}(t) \\
    &=\overline{\eta}(t) \, \overline{u}(t) \, .
  \end{aligned}
  \label{eq.cmplx_freq}
\end{equation}

The quantity $\overline{\eta}(t)$ is called \textit{complex frequency} of the variable $\overline{u}(t)$, and $\rho(t)$ and $\omega(t)$ are its real and imaginary parts, respectively.

\section{Dynamics of Park's Vectors}
\label{sec:dyn_park}
Park's vectors are time dependent complex quantities derived from the
$\rm dq$-axis components of the $\rm dqo$ transform, for example:
\begin{equation}
  \overline{{\imath}}(t) = \imath_{\mathrm{d}}(t)+ \jmath \, \imath_{\mathrm{q}}(t) \, ,
\end{equation}
where $\imath_{\mathrm{d}}$ and $\imath_{\mathrm{q}}$ can be obtained
from applying the $\rm dqo$-transform to the currents in $\rm abc$
coordinates, where we can drop the zero-sequence component by assuming
a balanced system.

To explore the dynamics of a Park's vector, we study its derivative
respect to time.  Accordingly, it is necessary to comprehend the
derivative in terms of the dq frame, which is thoroughly covered in
reference \cite{freqvar}.  In the Park reference frame, the time
derivative can be defined as the following operator:
\begin{equation}
    \overline{\mathrm{p}}(t) = \frac{d}{dt} + \jmath \, \omega_{\mathrm{dq}}(t) \, ,
    \label{eq.dq_der}
\end{equation}
where $\omega_{\mathrm{dq}}$ refers to the Park's reference angular frequency.

\section{Dynamics of the Current Flow}
\label{sec:dyn_curr}
We begin with the study of the relationship between the voltages of adjacent buses and the current flowing through a constant impedance $\overline{{Y}}_{hk}$ from node $h$ to $k$:
\begin{equation}
  \overline{{\imath}}_{hk} = \overline{{Y}}_{hk} 
  ( \overline{{v}}_h- \overline{{v}}_k) \, .
  \label{eq.iYv}
\end{equation}

Note that, in this work, the notation $\overline{{\imath}}_{hk}$
represents the specific current flowing from bus $h$ to $k$. On the
other hand, a single subindex, say $\overline{{\imath}}_{h}$,
represents the total injected current at bus $h$.  It is important to
note that within this equation, certain quantities are assumed to be
time-dependent.  Specifically, the voltages and current vectors
$\overline{{v}}_h$, $\overline{{v}}_k$, and $\overline{{\imath}}$ are
interpreted as dynamic variables, such as Park's vectors.

Assuming a constant admittance and using the derivative operator in (\ref{eq.dq_der}), we can properly derive (\ref{eq.iYv}) in time, as follows:
\begin{equation}
  \frac{d}{dt}\overline{\imath} + \jmath \, \omega_{\mathrm{dq}}
  \overline{\imath}= \overline{Y}_{hk} (
  \frac{d}{dt}\overline{v}_h+\jmath \,\omega_{\mathrm{dq}}
  \overline{v}_h- \frac{d}{dt}\overline{v}_k - \jmath \,
  \omega_{\mathrm{dq}} \overline{v}_k) \, ,
  \label{eq.diYv}
\end{equation}
where time dependence is not shown for simplicity.  We use this expression to represent the dynamics for the current through an ac branch as a function of the adjacent voltages.

\section{Performance index $\mu$}
\label{sec:mu}

The performance index $\mu$ was originally defined in \cite{derfv} based on the concept of the total variations of the voltage at a node. It is calculated to account for the combined effects of voltage magnitude and phase. This index is obtained by integrating the magnitude of the CF at a given bus $r$:
\begin{equation}
  \label{eq.mu}
  \mu_r(t) = \int_0^t | \bar{\eta}_r(\tau) | d\tau =
  \int_0^t \sqrt{\rho_r^2(\tau)+(\omega_r(\tau)-\omega_o)^2} \, d\tau \, .
\end{equation}
This index represents the normalized total variation of voltage over a time interval. The term $\omega_r-\omega_o$ is used to account for any deviation from $\overline{\eta}^{\mathrm{ref}}$, the steady-state condition for the CF. As $\mu_r$ is a cumulative metric, a smaller value indicates a more effective controller.

\footnotesize

\vspace{3mm}

\noindent \textbf{Rodrigo Bernal} received his BSc.~and MSc.~degrees in Electrical Engineering from Pontificia Universidad Cat\'olica de Chile in 2016 and 2020, respectively.  He has been actively involved in the field of power systems, working as a power system engineer in the industry since 2017.  Currently pursuing a Ph.D. in Electrical Engineering at University College Dublin, Ireland. His research interests focuses on power system modeling, control, and stability analysis.

\vspace{3mm}

\noindent \textbf{Federico Milano} (F’16) received from the Univ. of Genoa, Italy, the Ph.D.~in Electrical Engineering in 2003.  In 2013, he joined the University College Dublin, Ireland, where he is currently a full professor.  He is Secretary of the IEEE Power System Dynamic Performance Committee and past Chair of the Power System Stability Controls Subcommittee, IET Fellow, IEEE PES Distinguished Lecturer, Chair of the Technical Program Committee of the PSCC 2024, Senior Editor of the IEEE Transactions on Power Systems, Member of the Cigr\'e Irish National Committee, and Co-Editor in Chief of the IET Generation, Transmission \& Distribution.  His research interests include power system modeling, control and stability analysis.


\begin{thebibliography}{00}
\providecommand{\url}[1]{#1}
\csname url@samestyle\endcsname
\providecommand{\newblock}{\relax}
\providecommand{\bibinfo}[2]{#2}
\providecommand{\BIBentrySTDinterwordspacing}{\spaceskip=0pt\relax}
\providecommand{\BIBentryALTinterwordstretchfactor}{4}
\providecommand{\BIBentryALTinterwordspacing}{\spaceskip=\fontdimen2\font plus
\BIBentryALTinterwordstretchfactor\fontdimen3\font minus
  \fontdimen4\font\relax}
\providecommand{\BIBforeignlanguage}[2]{{%
\expandafter\ifx\csname l@#1\endcsname\relax
\typeout{** WARNING: IEEEtran.bst: No hyphenation pattern has been}%
\typeout{** loaded for the language `#1'. Using the pattern for}%
\typeout{** the default language instead.}%
\else
\language=\csname l@#1\endcsname
\fi
#2}}
\providecommand{\BIBdecl}{\relax}
\BIBdecl

\bibitem{converter_sync}
N.~Jaalam, N.~Rahim, A.~Bakar, C.~Tan, and A.~M. Haidar, ``A comprehensive
  review of synchronization methods for grid-connected converters of renewable
  energy source,'' \emph{Renewable and Sustainable Energy Reviews}, vol.~59,
  pp. 1471--1481, 2016.

\bibitem{converter_requirements_ctrlModes}
A.~Q. Al-Shetwi, M.~Hannan, K.~P. Jern, M.~Mansur, and T.~Mahlia,
  ``Grid-connected renewable energy sources: Review of the recent integration
  requirements and control methods,'' \emph{J.~of Cleaner Production}, vol.
  253, p. 119831, 2020.

\bibitem{Fundations}
F.~Milano, F.~Dörfler, G.~Hug, D.~J. Hill, and G.~Verbi\v{c}, ``Foundations
  and challenges of low-inertia systems (invited paper),'' in \emph{2018 Power
  Systems Computation Conference (PSCC)}, 2018, pp. 1--25.

\bibitem{GridForming2024} B.~Bahrani, M.~H. Ravanji, B.~Kroposki, D.~Ramasubramanian, X.~Guillaud, T.~Prevost, and N.-A.~Cutululis, ``Grid-Forming Inverter-Based Resource Research Landscape: Understanding the Key Assets for Renewable-Rich Power Systems,'' \emph{IEEE Power and Energy Magazine}, vol.~22, no.~2, pp.~18--29, 2024, doi: 10.1109/MPE.2023.3343338.

\bibitem{TimGreen_Duality}
Y.~Li, Y.~Gu, and T.~C. Green, ``Revisiting grid-forming and grid-following
  inverters: A duality theory,'' \emph{IEEE Trans.~on Power Systems}, vol.~37,
  no.~6, pp. 4541--4554, 2022.

\bibitem{osti_1813971}
Y.~Lin \emph{et~al.}, ``Stabilizing the power system in 2035 and beyond:
  Evolving from grid-following to grid-forming distributed inverter controllers
  (final technical report),'' 8 2021.

\bibitem{Sijia2022Unified}
S.~Geng and I.~A. Hiskens, ``Unified grid-forming/following inverter control,''
  \emph{IEEE OA J. of Power and Energy}, vol.~9, pp. 489--500, 2022.

\bibitem{ancserv}
C.~Demoulias \emph{et~al.}, ``Ancillary services offered by distributed
  renewable energy sources at the distribution grid level: An attempt at proper
  definition and quantification,'' \emph{App. Sciences}, vol.~10, no.~20, 2020.

\bibitem{PQDecoupling}
Y.~Laba, A.~Bruy{\`e}re, F.~Colas, and X.~Guillaud, ``{PQ} decoupling on
  grid-forming converter connected to a distribution network,'' in \emph{2022
  IEEE 13th International Symposium on Power Electronics for Distributed
  Generation Systems (PEDG)}, 2022, pp. 1--6.

\bibitem{TOULAROUD2023Hier}
M.~S. Toularoud, M.~K. Rudposhti, S.~Bagheri, and A.~H. Salemi, ``A
  hierarchical control approach to improve the voltage and frequency stability
  for hybrid microgrids-based distributed energy resources,'' \emph{Energy
  Reports}, vol.~10, pp. 2693--2709, 2023.

\bibitem{Wei2023Suple}
W.~Du, K.~P. Schneider, G.~P. Wiegand, F.~K. Tuffner, J.~Xie, and O.~L. Dent,
  ``A supplemental control for dynamic voltage restorers to improve the primary
  frequency response of microgrids,'' \emph{IEEE Trans.~on Smart Grid},
  vol.~14, no.~2, pp. 878--888, 2023.

\bibitem{RAJAN2021PFC}
R.~Rajan, F.~M. Fernandez, and Y.~Yang, ``Primary frequency control techniques
  for large-scale pv-integrated power systems: A review,'' \emph{Renewable and
  Sustainable Energy Reviews}, vol. 144, p. 110998, 2021.

\bibitem{OptimalDroop}
B.~Alghamdi and C.~A. Cañizares, ``Frequency regulation in isolated microgrids
  through optimal droop gain and voltage control,'' \emph{IEEE Trans.~on Smart
  Grid}, vol.~12, no.~2, pp. 988--998, 2021.

\bibitem{Adaptive_Inertia_Control}
L.~Huang, C.~Yang, M.~Song, H.~Yuan, H.~Xie, H.~Xin, and Z.~Wang, ``An adaptive
  inertia control to improve stability of virtual synchronous machines under
  various power grid strength,'' in \emph{IEEE PES General Meeting}, 2019, pp.
  1--5.

\bibitem{BandwidthSep}
Y.~Zhang, Y.~Li, Y.~Gu, and T.~C. Green, ``Consideration of control-loop
  interaction in transient stability of grid-following inverters using
  bandwidth separation method,'' in \emph{2023 11th International Conference on
  Power Electronics and ECCE Asia (ICPE 2023 - ECCE Asia)}, 2023, pp.
  1294--1301.

\bibitem{symcontrol_coupling}
X.~Zhang, S.~Fu, W.~Chen, N.~Zhao, G.~Wang, and D.~Xu, ``A symmetrical control
  method for grid-connected converters to suppress the frequency coupling under
  weak grid conditions,'' \emph{IEEE Trans.~on Power Electronics}, vol.~35,
  no.~12, pp. 13\,488--13\,499, 2020.
  
\bibitem{FAZAL2023}
S.~Fazal, M.~{Enamul Haque}, M.~{Taufiqul Arif}, A.~Gargoom, and A.~M.~T. Oo,
  ``Grid integration impacts and control strategies for renewable based
  microgrid,'' \emph{Sustainable Energy Technologies and Assessments}, vol.~56,
  p. 103069, 2023.

\bibitem{freqvoltage}
G.~Tzounas and F.~Milano, ``Improving the frequency response of {DERs} through
  voltage feedback,'' in \emph{IEEE PES General Meeting}, 2021, pp. 1--5.

\bibitem{VCFC_Impedance}
K.~De~Brabandere, B.~Bolsens, J.~Van~den Keybus, A.~Woyte, J.~Driesen, and
  R.~Belmans, ``A voltage and frequency droop control method for parallel
  inverters,'' \emph{IEEE Trans.~on Power Electronics}, vol.~22, no.~4, pp.
  1107--1115, 2007.

\bibitem{bernal2024improving}
R.~Bernal and F.~Milano, ``Improving voltage and frequency control of {DERs}
  through dynamic power compensation,'' in \emph{Proceedings of the 23rd Power
  System Computation Conference (PSCC)}, Paris-Saclay, France, June 4-7 2024,
  accepted for presentation.

\bibitem{Farrokhabadi2017}
M.~Farrokhabadi, C.~A. Cañizares, and K.~Bhattacharya, ``Frequency control in
  isolated/islanded microgrids through voltage regulation,'' \emph{IEEE
  Trans.~on Smart Grid}, vol.~8, no.~3, pp. 1185--1194, 2017.

\bibitem{derfv}
W.~Zhong, G.~Tzounas, and F.~Milano, ``Improving the power system dynamic
  response through a combined voltage-frequency control of distributed energy
  resources,'' \emph{IEEE Trans.~on Power Systems}, vol.~37, no.~6, pp.
  4375--4384, 2022.

\bibitem{FastVoltageBoosters}
R.~E. {\'A}vila-Mart{\'i}nez \emph{et~al.}, ``Fast voltage boosters to improve
  transient stability of power systems with 100\% of grid-forming {VSC}-based
  generation,'' \emph{IEEE Trans.~on Energy Conversion}, vol.~37, no.~4, pp.
  2777--2789, 2022.

\bibitem{curvatureCtrl}
F.~Sanniti, G.~Tzounas, R.~Benato, and F.~Milano, ``Curvature-based control for
  low-inertia systems,'' \emph{IEEE Trans.~on Power Systems}, vol.~37, no.~5,
  pp. 4149--4152, 2022.

\bibitem{cmplx}
F.~Milano, ``Complex frequency,'' \emph{IEEE Trans.~on Power Systems}, vol.~37,
  no.~2, pp. 1230--1240, 2022.

\bibitem{moutevelis2023taxonomy}
D.~Moutevelis, J.~Rold{\'a}n-P{\'e}rez, M.~Prodanovic, and F.~Milano,
  ``Taxonomy of power converter control schemes based on the complex frequency
  concept,'' 2023.

\bibitem{buttner2023}
A.~B\"uttner and F.~Hellmann, ``Complex couplings -- {A} universal, adaptive
  and bilinear formulation of power grid dynamics,'' \emph{arXiv}, no.
  2308.15285, 2023.

\bibitem{complexfrequencysync}
X.~He, V.~Häberle, and F.~Dörfler, ``Complex-frequency synchronization of
  converter-based power systems,'' 2022.

\bibitem{ConverterTech}
Y.~Wang, O.~Lucia, Z.~Zhang, S.~Gao, Y.~Guan, and D.~Xu, ``A review of high
  frequency power converters and related technologies,'' \emph{IEEE Open J. of
  the Industrial Electronics Society}, vol.~1, pp. 247--260, 2020.

\bibitem{GF_converters}
R.~Rosso, X.~Wang, M.~Liserre, X.~Lu, and S.~Engelken, ``Grid-forming
  converters: Control approaches, grid-synchronization, and future trends—a
  review,'' \emph{IEEE Open J. of Ind. Apps.}, vol.~2, pp. 93--109, 2021.

\bibitem{circlelim}
B.~Fan and X.~Wang, ``Equivalent circuit model of grid-forming converters with
  circular current limiter for transient stability analysis,'' \emph{IEEE
  Trans.~on Power Systems}, vol.~37, no.~4, pp. 3141--3144, 2022.

\bibitem{awulim}
J.~Chen, C.~Ge, D.~Qiang, H.~Geng, T.~O'Donnell, and F.~Milano, ``Impact of
  frequency anti-windup limiter on synchronization stability of grid feeding
  converter,'' \emph{CSEE J.~of Power and Energy Systems}, vol.~9, no.~5, pp.
  1676--1687, 2023.

\bibitem{Sauer_Book}
P.~Sauer and M.~Pai, \emph{\BIBforeignlanguage{English (US)}{Power system
  dynamics and stability}}.\hskip 1em plus 0.5em minus 0.4em\relax Prentice
  Hall, 1998.

\bibitem{REGFM_A1}
\BIBentryALTinterwordspacing
W.~Du, ``Model specification of droop-controlled, grid-forming inverters
  (regfm\_a1),'' 9 2023. [Online]. Available:
  \url{https://www.osti.gov/biblio/2229442}
\BIBentrySTDinterwordspacing

\bibitem{dome}
F.~Milano, ``A {Python}-based software tool for power system analysis,'' in
  \emph{IEEE PES General Meeting}, 2013, pp. 1--5.

\bibitem{PLL}
A.~Ortega and F.~Milano, ``Comparison of different pll implementations for
  frequency estimation and control,'' in \emph{2018 18th International
  Conference on Harmonics and Quality of Power (ICHQP)}, 2018, pp. 1--6.

\bibitem{PQStc}
M.~Adeen and F.~Milano, ``On the impact of data-driven stochastic load models
  on power system dynamics,'' in \emph{IEEE PES General Meeting}, 2023, pp.
  1--5.

\bibitem{IRISH}
M.~Adeen, F.~Bizzarri, D.~d. Giudice, S.~Grillo, D.~Linaro, A.~Brambilla, and
  F.~Milano, ``On the calculation of the variance of algebraic variables in
  power system dynamic models with stochastic processes,'' \emph{IEEE Trans.~on
  Power Systems}, vol.~38, no.~2, pp. 1739--1742, 2023.

\bibitem{freqvar}
F.~Milano and {\'A}.~Ortega~Manjavacas, \emph{Frequency Variations in Power
  Systems}.\hskip 1em plus 0.5em minus 0.4em\relax Wiley-IEEE Press, June 2020.

\bibitem{eirgrid2024gridcode}
EirGrid, \emph{Grid Code Version 14.2}, Jul. 2024. [Online]. Available: \url{https://cms.eirgrid.ie/sites/default/files/2024-07/GridCodeVersion14.2docx.pdf}. 

\end{thebibliography}
\end{document}